\documentclass[french]{article-hermes}
\usepackage{graphicx}

\newcommand{\julia}{\textsc{Julia}}
\newcommand{\fractal}{\textsc{Fractal}}
\begin{document}

\title[Reconfiguration d'implantation]{Support pour la reconfiguration d'implantation dans les applications à composants Java}

\author{Jakub Korna\'s \andauthor Matthieu Leclercq \andauthor Vivien Qu\'ema\\Jean-Bernard Stefani}

\address{Laboratoire LSR-IMAG (CNRS, INPG, UJF) - INRIA - projet Sardes\\
INRIA Rhône-Alpes, 655 av. de l'Europe, F-38334 Saint-Ismier cedex\\
Prenom.Nom@inrialpes.fr}

\resume{
De nombreux modèles de composants sont aujourd'hui utilisés à des fins variées~: construction d'applications, d'intergiciels, ou encore de systèmes d'exploitation. Ces modèles permettent tous des reconfigurations de structure, c'est-à-dire des modifications de l'architecture de l'application. En revanche, peu permettent des reconfigurations d'implantation qui consistent à modifier dynamiquement le code des composants de l'application. Dans cet article nous présentons le travail que nous avons effectué dans {\julia}, une implémentation Java du modèle {\fractal}, pour permettre le dynamisme d'implantation. Nous montrons comment les limitations du mécanisme de chargement de classes Java ont été contournées pour permettre de modifier les classes d'implémentation et d'interfaces des composants. Nous décrivons également l'intégration de notre proposition avec l'ADL de {\julia}.
}

\abstract{Nowadays, numerous component models are used for various purposes: to build
applications, middleware or even operating systems. Those models
commonly support structure reconfiguration, that is modification of
application's architecture at runtime. On the other hand, very few allow
implementation reconfiguration, that is runtime modification of the code
of components building the application. In this article we present the
work we performed on {\julia}, a Java-based implementation of the {\fractal}
component model, in order for it to support implementation
reconfigurations. We show how we overcame the limitations of Java class
loading mechanism to allow runtime modifications of components'
implementation and interfaces. We also describe the integration of our
solution with the {\julia} ADL.
}

\motscles{Systèmes à composants, Java, Reconfiguration dynamique, Class loader, Fractal}


\proceedings{DECOR'04, Déploiement et (Re)Configuration de Logiciels}{171}

\maketitlepage

\section{Introduction}

Les modèles de composants ont fait leur apparition dans les deux dernières décennies. Ils sont désormais utilisés pour construire de nombreux systèmes, aussi bien au niveau applicatif (EJB~\cite{ejb02}, CCM~\cite{ccm01}) qu'au niveau intergiciel (dynamicTAO~\cite{kon01dynamic}, OpenORB~\cite{blair01design}), où encore au niveau système (OSKit~\cite{Ford97:OSKit}, THINK~\cite{fassino02think}). Un des apports des composants est qu'ils sont des briques logicielles indépendantes pouvant être assemblées dynamiquement pour construire des logiciels complexes. L'aspect dynamique de l'assemblage prend une importance croissante du fait de l'évolution très rapide des équipements informatiques et des besoins des utilisateurs d'applications.


Il est aujourd'hui communément admis de distinguer deux formes de reconfiguration dynamique d'applications~\cite{hofmeister94dynamic} : {\it reconfiguration de structure} et {\it reconfiguration d'implantation}. Dans un système à composants, une reconfiguration de structure consiste à ajouter/enlever un composant, modifier une liaison entre composants, ou encore modifier la localisation d'un composant (si l'application est distribuée). Une reconfiguration d'implantation consiste à mettre à jour le code d'un composant ou d'une ou plusieurs de ses interfaces.

Si les modèles de composants permettent toujours des reconfigurations de structure, il n'en n'est pas de même pour les reconfigurations d'implantation, dont la mise en {\oe}uvre peut être très difficile --- voire impossible --- suivant le langage d'implantation du modèle de composants. Dans cet article, nous nous focalisons sur les modèles de composants implantés en Java. Java offre un système dynamique de chargement de classes, appelé {\it class loader}. Celui-ci impose cependant un certain nombre de contraintes d'utilisation qui rendent difficile la mise à jour de code. Il est par exemple impossible de charger deux versions d'une même classe, ou encore de décharger une classe. Deux possibilités s'offrent aux développeurs Java pour fournir des mécanismes de reconfiguration d'implantation : (1) modification du code des classes chargées pour garantir que deux versions d'une même classe portent des noms différents ; (2) utilisation de class loaders différents pour charger les différentes versions d'une même classe. La première solution préclut l'utilisation d'une partie du langage Java : il n'est, par exemple, plus possible d'utiliser la méthode de construction de classe {\tt Class.forName(``toto'')}, étant donné que celle-ci se base sur le nom de la classe à créer --- nom qui peut avoir été modifié lors du chargement. La seconde solution est utilisée dans plusieurs serveurs J2EE~: ceux-ci utilisent des hiérarchies en arbre de class loaders, chaque class loader ayant un unique parent auquel il peut éventuellement déléguer le chargement de classes. Ces organisations en arbre ne sont pas assez flexibles, et ne permettent que très peu de reconfigurations d'implantation.      

Dans cet article, nous présentons le travail que nous avons mené au sein de {\julia}~\cite{bruneton04open}, une implantation Java du modèle de composants {\fractal}, pour permettre la co-existence de multiples versions d'une même classe, et donc la reconfiguration d'implantation d'une application à composants. La solution que nous adoptons est d'utiliser une organisation arbitraire de class loaders : cette organisation est construire en se basant sur les frontières de composants et les besoins de reconfigurabilité exprimés par le développeur d'applications à l'aide du langage de description d'architecture de {\julia}. Les class loaders sont créés et administrés à l'aide de {\it Module loader}~\cite{hall04policy}, un système permettant de faire coopérer un ensemble de class loaders.
  
L'article s'organise de la façon suivante : nous présentons le modèle de composants {\fractal} et son implantation, {\julia}, dans la section~\ref{sec:fractal}. La section~\ref{sec:proposition} présente notre proposition pour la reconfiguration d'implantation dans {\julia}. Son implémantation est décrite dans la section~\ref{sec:implementation}. Enfin, nous présentons les travaux connexes dans la section~\ref{sec:travaux_connexes}, avant de conclure cet article.

\section{Le modèle de composants {\fractal}}\label{sec:fractal}
\begin{sloppypar}
Le modèle de composants {\fractal}~\cite{bruneton04fractal,bruneton04open} est un modèle flexible et réflexif. Contrairement à d'autres modèles comme Jiazzi~\cite{McDirmid01:Jiazzi} ou ArchJava~\cite{Aldrich02:ArchJava}, {\fractal} n'est pas une extension à un langage, mais une librairie qui permet la création et la manipulation de composants et d'architectures à base de composants. Dans la suite de cette section, nous présentons {\julia}, une implantation Java de {\fractal}
\end{sloppypar}

\subsection{Le modèle}

{\julia} distingue deux types de composants : les composants {\it primitifs} sont essentiellement des classes Java standards avec quelques conventions de codage. Les composants {\it composites} encapsulent un groupe de composants primitifs et/ou composites. Une caractéristique originale du modèle est qu'un composant peut être encapsulé simultanément dans plusieurs composites. Un tel composant est appelé composant {\it partagé}.   

Un composant est constitué de deux parties : la partie de {\it contrôle} --- qui expose les interfaces du composant et comporte des objets contrôleurs et intercepteurs ---, et la partie {\it fonctionnelle} --- qui peut être soit une classe Java (pour les composants primitifs), soit des sous-composants (pour les composants composites). 

Les points d'accès d'un composant sont appelés {\it interfaces}. On distingue deux types d'interfaces : les interfaces serveurs sont des points d'accès acceptant des appels de méthodes entrants. Elles correspondent donc aux services fournis par le composant. Les interfaces clients sont des points d'accès permettant des appels de méthodes sortants. Elles correspondent aux services requis par le composant. Ces deux types d'interfaces sont décrits par une signature Java standard et une indication sur le rôle de l'interface (client ou serveur). 

La communication entre composants {\julia} est uniquement possible si leurs interfaces sont liées. {\julia} supporte deux types de liaisons : {\it primitives} et {\it composites}. Une liaison primitive est une liaison entre une interface cliente et une interface serveur appartenant au même espace d'adressage. Elle signifie que les appels de méthodes émis par l'interface cliente doivent être acceptés par l'interface serveur. Une telle liaison est dite ``primitive'' car elle est mise en {\oe}uvre de façon très simple à l'aide d'une référence Java. Une liaison composite est un chemin de communication entre un nombre arbitraire d'interfaces de composants. Ces liaisons sont construites à l'aide d'un ensemble de liaisons primitives et de composants de liaisons (stubs, skeletons, adaptateurs, etc). 

\subsection{Exemple}

\begin{figure}[!htbp]
\centering
\includegraphics[scale=0.6]{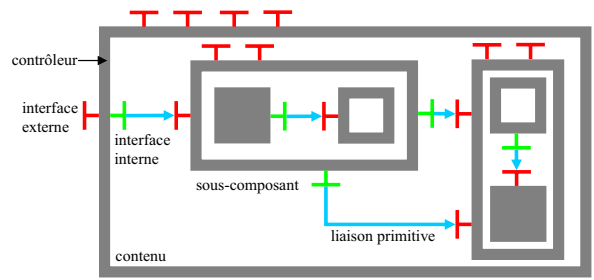}
\caption{Exemple d'application {\fractal}}
\label{fig:fractal_component}
\end{figure}

La figure~\ref{fig:fractal_component} illustre les différents concepts du modèle de composants {\fractal}. Les rectangles représentent des composants: la partie grisée correspond à la partie {\it contrôle} des composants~; l'intérieur du rectangle correpond au contenu des composants. Les flêches correspondent aux liaisons entre interfaces (représentées par des structures en forme de ``T''). Les interfaces externes apparaissant au sommet des composants sont les interfaces de contrôle. Les deux rectangles gris représentent un composant partagé.

\subsection{Implantation}
Un composant {\julia} est formé de plusieurs objets Java que l'on peut séparer en trois groupes (figure~\ref{fig:julia_structures}): 

\begin{figure}[!htbp]
\centering
\includegraphics[scale=0.38]{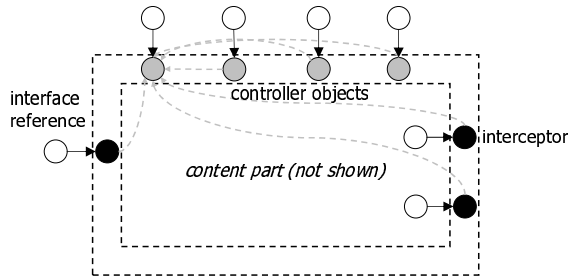}
\caption{Implantation d'un composant {\julia}}
\label{fig:julia_structures}
\end{figure}

\begin{itemize}

\item Les objets qui implémentent le contenu du composant. Ces objets n'ont pas été représentés sur la figure. Ils peuvent être des sous-composants (dans le cas de composants composites) ou des objets Java (pour les composants primitifs).

\item Les objets qui implémentent la partie de contrôle du composant (représentés en noir et en gris). Ces objets peuvent être séparés en deux groupes : les objets implémentant les interfaces de contrôle, et des intercepteurs optionnels qui interceptent les appels de méthodes entrant et sortant sur les interfaces fonctionnelles. 

\item Les objets qui référencent les interfaces du composant (en blanc). 
\end{itemize}

La mise en place de ces différents objets est effectuée par des fabriques de composants. Celles-ci fournissent une méthode de création qui prend en paramètres une description des parties fonctionnelles et de contrôle du composant. Dans l'implémentation actuelle de {\julia}, toutes les fabriques de composants utilisent le même class loader. 

\section{Vers une reconfiguration d'implantation dans {\julia}}\label{sec:proposition}

Tel que défini dans la section précédente, {\julia} fournit un support pour la reconfiguration de structure : en effet, les contrôleurs des composants permettent d'ajouter/retrancher des composants, ou encore de modifier les liaisons entre composants. En revanche, {\julia} ne fournit aucun support pour la reconfiguration d'implantation. En effet, l'ensemble des classes chargées dans une machine virtuelle Java (i.e. dans le même espace d'adressage) sont chargées par le même class loader. En conséquence, il n'est pas possible de faire co-exister deux versions d'une même classe, ce qui exclut les reconfigurations d'implantation. Nous commençons cette section par une description des différentes contraintes imposées par le class loader Java, puis nous décrivons notre proposition.

\subsection{Les contraintes du class loader Java}

Le class loader Java~\cite{liang98dynamic} permet de charger dynamiquement des classes Java au sein d'une machine virtuelle. Chaque class loader charge les classes à partir d'une source définie : système de fichiers, réseau, etc. Le rôle d'un class loader est de créer des objets {\tt Class} à partir de fichiers {\tt .class}. Pour ce faire, chaque class loader implémente une méthode {\tt loadClass(String name)} qui a pour but de charger une classe. Cette méthode utilise la méthode {\tt findLoadedClass(String name)} pour vérifier que la classe n'a pas déjà  été chargée, auquel cas elle est stockée dans un cache. Si tel n'est pas le cas, {\tt loadClass} utilise {\tt defineClass} pour construire un objet {\tt Class} à partir d'un tableau d'octets. Notons qu'un class loader peut également déléguer le chargement d'une classe à un autre class loader.  Ce mécanisme est majoritairement utilisé avec des hiérarchies en arbre de class loaders : chaque class loader a un père auquel il peut déléguer le chargement. 

Il est important de noter qu'une classe est liée au class loader qui l'a chargée : la même classe chargée par deux class loaders différents est considérée comme deux classes différentes. En conséquence, l'utilisation d'une classe à la place de l'autre lève une exception {\tt ClassCastException}.
 

\subsection{Les class loaders dans un monde de composants}



Une application {\fractal} est constituée d'un ensemble de composants. Chaque composant est constitué d'un ensemble de classes (décrites dans la section~\ref{sec:fractal}). La question à laquelle nous devons répondre est~: à quels class loaders faut-il déléguer le chargement de chacune des classes~? 


Selon Szyperski~\cite{szyperski98component}, ``un composant est déployable indépendamment et est sujet à composition par une tierce partie''. Une approche naïve pour rendre les composants deployables indépendamment consiste à charger chaque composant dans un class loader indépendant. Cependant de tels composants ne sont pas composables par une tierce partie. En effet, toute tentative de liaison entre les interfaces de deux composants résulterait en une exception {\tt ClassCastException}. 

Il apparaît donc logique de séparer les {\it classes d'interfaces} du composant (fonctionnelles et de contrôle) --- destinées à être utilisées par les autres composants ---, des {\it classes d'implémentation} --- destinées à l'usage ``privé'' du composant. Supposons que l'on ait deux composants $C_1$ et $C_2$ qui communiquent via une interface {\tt Push} qui définit une méthode {\tt void push(Message m)}. Les classes {\tt Push} et {\tt Message} doivent être chargées par un class loader commun aux deux composants tandis que les classes d'implantation des composants peuvent être chargées indépendamment. 

Cette solution n'est néanmoins pas suffisante. Supposons que la signature de la méthode {\tt push} soit différente~: {\tt void push(Object o)}. Il est nécessaire que toutes les classes des objets transitant via la méthode {\tt push} soient chargées dans un class loader commun à $C_1$ et $C_2$. Ces classes, appelées {\it classes partagées}, sont un sous-ensemble des classes d'implémentation des composants. 

\section{Mise en {\oe}uvre}\label{sec:implementation}

Dans cette section, nous présentons la mise en {\oe}uvre de la proposition formulée dans la section précédente. Nous décrivons l'utilisation de {\it Module loader} pour l'intégration des classes loaders dans {\julia}, puis nous montrons comment l'ADL de {\julia} a été étendu pour permettre la spécification de versions d'interfaces, du code partagé, et l'instanciation des différents class loaders. 


\subsection{Module loader}

{\it Module loader}~\cite{hall04policy} est un canevas permettant de construire des organisations arbitraires de class loaders, respectant une logique de chargement définie par le développeur. Les principaux concepts de {\it Module loader} sont~: {\it module}, {\it module manager}, {\it resource source} et {\it search policy}.

\begin{itemize}
\item Le {\bf module} est une unité logique de groupement de ressources (i.e classes). Chaque module est associé à un class loader. Par ailleurs, chaque module peut définir des méta-données. 

\item Un {\bf module manager} gère un ensemble de modules. Il envoie des notifications quand des modules sont ajoutés ou retranchés.

\item Une {\bf resource source} est une source à partir de laquelle un module peut charger des classes.  Cette source peut être un fichier, une URL, une base de données, etc. Les concepteurs de modules peuvent définir leurs propres sources.

\item La {\bf search policy} est le ``cerveau'' du mécanisme de chargement de classes. Il définit la façon dont un module peut localiser (et donc charger) une classe. La {\it search policy} adapte son comportement en fonction des événements qu'elle reçoit du {\it module manager}. Un exemple typique de {\it search policy} est l'{\it import search policy}~: les modules annoncent (via des méta-données) les ressources qu'ils exportent (i.e. fournissent) aux autres modules, et les ressources qu'ils importent (i.e. requièrent) des autres modules. Un module ne peut être créé que si les ressources qu'il importe sont présentes.

\end{itemize}

\subsection{Utilisation de Module loader dans {\julia}}

\subsubsection{Les différentes structures de données}

Comme nous l'avons vu dans la section~\ref{sec:proposition}, chaque composant utilise des class loaders différents pour les classes d'implémentation, d'interfaces, et les classes partagées. L'information sur les classes requises par un composant est stockée dans un module, appelé {\it Info Module}. Un {\it Info Module} ne charge pas de classes. Il délègue le chargement des classes à des {\it Resource Modules}, qui donnent accès à des classes et à des informations sur ces classes. Les {\it Resource Modules} sont créés en accord avec les règles énoncées dans la section~\ref{sec:proposition}~: un module par implémentation de composant, un module par classe d'interface, et un module pour les classes partagées. 

\begin{figure}[!ht]
\centering
\includegraphics[scale=0.5]{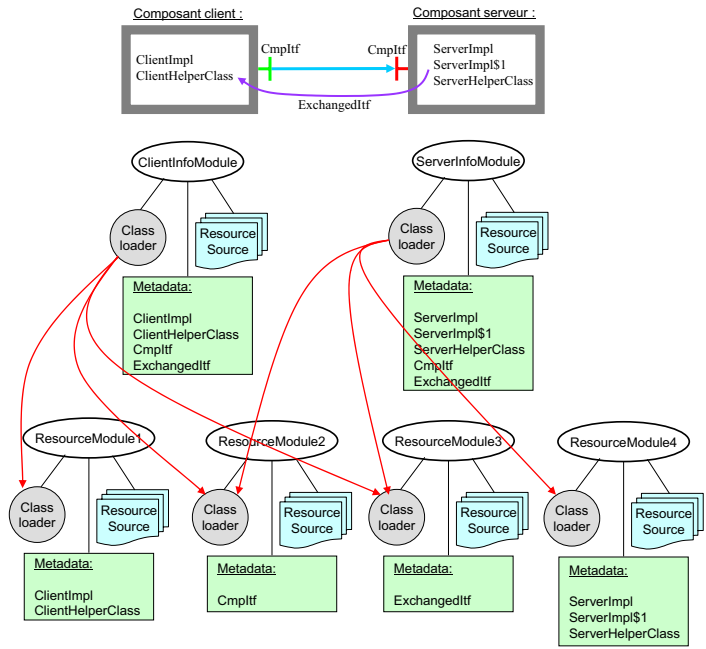}
\caption{Organisation des modules}
\label{fig:organisation_modules}
\end{figure}

La figure~\ref{fig:organisation_modules} présente un exemple d'organisation de modules. L'application fait intervenir deux composants, chacun d'eux étant implémenté par un certain nombre de classes. Les composants communiquent via une interface représentée par l'interface {\tt CmpItf}. Par ailleurs, les composants échangent une {\it classe partagée} {\tt ExchangedItf}. L'organisation des modules respecte les règles énoncées au paragraphe précédent~: chaque composant est associé à un {\it Info Module} qui délègue le chargement des classes à différent {\it Resource Modules}. Cette délégation se base sur les méta-données de chaque module. 

\subsubsection{Comportement des modules à l'exécution}

Quand un composant est créé, le class loader de son {\it Info Module} devient le class loader courant. Quand une classe doit être chargée, l'{\it Info Module} cherche un {\it Ressource Module}, parmi ceux qu'il connaît, qui peut charger la classe requise dans la version appropriée. Cette classe est chargée par le class loader associé au {\it Ressource Module}. 

Par ailleurs, il est nécessaire que lorsque les composants communiquent, le flot d'exécution des requêtes (i.e le {\it thread}) soit associé aux class loaders appropriés. Par exemple, si un appel ``traverse'' quatre composants, il est nécessaire que lors du transit de l'appel dans chacun des composants, le class loader du flot d'exécution soit mis à jour~: pour chaque composant, il doit être positionné sur l'{\it Info Module} du composant. Pour ce faire, nous avons développé un intercepteur dont le rôle est d'intercepter toutes les requêtes pour modifier le class loader du flot d'exécution. Le mécanisme d'interception engendre un léger surcoût des temps d'exécution. Ce surcoût est de l'ordre de 3 à 4\%.

\subsection{Extension de l'ADL {\julia}}

Le langage de description d'architecture (ADL) de {\julia} permet de décrire des applications à base de composants {\julia}. L'ADL a été conçu de manière extensible~: il est composé de plusieurs modules, chacun définissant une syntaxe pour exprimer un ``aspect'' de l'architecture (interfaces, liaisons, attributs, etc.). Les développeurs sont libres de créer leurs propres modules. 

Une fabrique est en charge d'utiliser la description de l'application pour créer l'architecture correspondante (création et liaison des composants, positionnement des attributs, etc.). De façon similaire à l'ADL, cette fabrique a été conçue de manière extensible, afin de pouvoir y intégrer du code relatif aux modules ADL définis par le développeur. 

Nous avons créé un module permettant de spécifier les versions de composants et d'interfaces utilisées par l'application. Ce module permet également de spécifier les {\it classes partagées} de chacun des composants. La figure~\ref{fig:exemple_adl} donne un exemple de description ADL. L'attribut {\tt version} a été ajouté aux définitions d'interfaces et de contenu. Par ailleurs, l'élément {\tt file} permet de spécifier les classes qui sont partagées, i.e. qui sont échangées par les composants, via leurs interfaces. 

\begin{figure}[!htbp]
\centering
\begin{small}
\begin{tabular}{|p{9.8cm}|}
\hline       
\begin{verbatim}
<definition name="HelloWorld" version="2.0">
    <interface name="r" role="server" 
               signature="java.lang.Runnable"/>
    <component name="client">
        <interface name="r" role="server" 
                   signature="java.lang.Runnable"/>
        <interface name="s" role="client" 
                   signature="Service" version="1.0"/>
        <content class="ClientImpl" version="1.0"/>
        <file name="Request" version="1.0"/>
    </component>
    <component name="server">
        <interface name="s" role="server" 
                   signature="Service" version="1.0"/>
        <content class="ServerImpl" version="2.0"/>
        <file name="Request" version="1.0"/>
    </component>
    <binding client="this.r" server="client.r"/>
    <binding client="client.s" server="server.s"/>
</definition>
\end{verbatim}
\\
\hline
\end{tabular}
\end{small}
\vspace*{6pt}
\caption{Un exemple de description ADL}
\label{fig:exemple_adl}
\end{figure}

Nous avons également développé un module pour la fabrique associée à l'ADL. Ce module a la responsabilité de déterminer les class loaders nécessaires, et de créer les {\it Info Modules} et {\it Ressources Modules} appropriés. Le développeur d'applications peut configurer deux aspects de ce module~: la granularité des class loader créés (en fonction de la granularité de reconfiguration souhaitée), et les sources des classes requises par l'application. A l'heure actuelle, nous n'offrons la possibilité que de deux granularités~: un seul class loader --- ce qui correspond au comportement de {\julia} et ne permet pas de reconfiguration dynamique ---, et la granularité définie dans les sections précédentes. Nous sommes en train de développer un module permettant au développeur de spécifier les composants qu'il veut pouvoir charger/décharger dynamiquement. Concernant la spécification des sources de classes, nous ne permettons actuellement que de spécifier des fichiers {\tt .class} et des fichiers {\tt.jar}.

\section{Travaux connexes}\label{sec:travaux_connexes}

On peut distinguer trois courants dans les travaux sur le dévelopement de canevas à composants extensibles en Java~: les modèles de composants, les plates-formes de services, et les serveurs J2EE.

Parmi les modèles de composants, citons JPloy~\cite{luer04jploy} et SOFA~\cite{hnetynka03fighting}. Ces deux modèles utilisent des manipulations de bytecode pour garantir que les différentes versions d'une même classe sont chargées avec des noms différents. Ces noms sont générés à partir de fichiers de descriptions (comparables à des descriptions ADL) qui spécifient les versions des classes utilisées. L'avantage de cette approche est qu'elle n'engendre aucun surcoût à l'exécution. En revanche, elle rend impossible l'utilisation de certaines méthodes du langage Java. Il n'est, par exemple, pas possible d'utiliser certaines méthodes de construction de classe par réflexion (e.g. {\tt Class.forName(String name)}).

Les plates-formes de services sont apparues ces dernières années avec les travaux menés sur OSGi~\cite{osgi03}. OSGi permet de déployer des applications Java empaquetées sous formes de {\it bundles}. Un bundle contient des fichiers jar et des méta-données sur ces fichiers jar (version, etc.). Le rôle d'une plate-forme OSGi est de gérer le cycle de vie des bundles (déploiement, activation, retrait, etc.). Un des apports d'OSGi pour la communauté Java est la prise en compte des versions des fichiers jar. Néanmoins, OSGi impose certaines contraintes~: il ne permet, par exemple, pas de faire co-exister deux versions d'une même classe. Par ailleurs, le modèle de composants utilisé dans OSGi est un modèle plat~: il n'est pas possible de créer des bundles ``composites''.

Enfin, de nombreux travaux sont effectués sur le chargement de classes dans les serveurs J2EE~\cite{j2ee02} (IBM WebSphere~\cite{webspherecl}, BEA WebLogic~\cite{weblogiccl}, JOnAS~\cite{jonas02cl}, etc.). Le but est de pouvoir isoler les différentes applications déployées sur un serveur. Les class loaders sont organisés en arbre, ce qui s'avère suffisant pour les applications ciblées, mais qui est trop restrictif dans un modèle de composants plus flexible comme {\fractal}.

\section{Conclusion}\label{sec:conclusion}

Il existe aujourd'hui de nombreux modèles de composants utilisés dans des domaines variés~: applications pour l'Internet, intergiciels, systèmes d'exploitation, etc. Si ces modèles supportent tous le dynamisme de structure --- qui consiste en une modification de l'architecture de l'application s'exécutant ---, peu, en revanche, supportent le dynamisme d'implantation --- qui consiste à modifier dynamiquement le code des composants de l'application. 

Dans cette article, nous avons présenté le travail que nous avons effectué au sein de {\julia}, une implémentation Java du modèle {\fractal}, pour ajouter des capacités de reconfiguration d'implantation. Nous avons montré comment contourner les limitations imposées par le class loader --- le mécanisme Java de chargement de code --- pour permettre aux composants de modifier leur implémentation et leurs interfaces.

La solution que nous proposons est plus flexible que celle mise en {\oe}uvre dans les serveurs J2EE, et ne présente pas les contraintes de celles reposant sur des modification du bytecode des classes des composants. Par ailleurs, l'intégration de notre proposition dans l'ADL {\julia} la rend transparente au développeur de composants.  

Nous poursuivons actuellement nos travaux pour permettre au développeur de spécifier le niveau de granularité qu'il souhaite, c'est-à-dire les composants qu'il souhaite pouvoir charger/décharger dynamiquement.

\bibliography{decor2004}

\end{document}